\documentstyle[11pt,newpasp,twoside, psfig, epsf]{article}
\def\edcomment#1{\iffalse\marginpar{\raggedright\sl#1\/}\else\relax\fi}
\reversemarginpar

\begin{document}
\title{The Duality of Spiral Structure, and a Quantitative Dust
  Penetrated Morphological Tuning Fork at Low and High Redshift}

\author{D. L. Block, I. Puerari and R. J. Buta}
\affil{Department of Computational and Applied Mathematics, University of the
Witwatersrand, Johannesburg, South Africa}
\author{R. Abraham}
\affil{Department of Astronomy, University of Toronto, Toronto, Canada}
\author{M. Takamiya}
\affil{Gemini Observatory, Hilo, HI}
\author{A. Stockton}
\affil{Institute for Astronomy, University of Hawaii, Honolulu, HI}

\begin{abstract}
In the near-infrared, the morphology of older star-dominated disks
indicates a simple classification scheme (1) H$m$ where $m$ is the
dominant harmonic, (2) a pitch angle (derived from the Fourier
spectra) associated with the rate of shear A/$\omega$ in the stellar
disk and (3) a `bar strength' parameter, robustly derived from the
gravitational potential or torque of the bar.  A spiral galaxy may
present two radically different morphologies in the optical and
near-infrared regime; there is no correlation between our quantitative
dust penetrated tuning fork and that of Hubble. Applications of our
$z\sim$0 Fourier template to the HDF are discussed using $L$ and $M$
band simulations from an 8-m NGST; the rest-wavelength IR morphology
of high-$z$ galaxies should probably be a key factor in deciding the
final choice of instruments for the NGST.
\end{abstract}

\section{Introduction}

{\it There is a fundamental limit in predicting what an evolved
stellar disk might look like} (Block et al. 1999). The greater the
degree of decoupling, the greater is the uncertainty.  The fact that a
spiral might be flocculent in the optical is very important, but it is
equally important to know whether or not driving the dynamics is a
grand design old stellar disk.

Decouplings between stellar and gaseous disks are cited in many
studies including Grosb{\o}l \& Patsis (1998), Elmegreen et al.
(1999), Block et al. (1999) and Puerari et al. (2000). The Hubble type
of a galaxy does not dictate its dynamical mass distribution (Burstein
\& Rubin 1985). {For example, {\it the Fourier spectra of the evolved
disks of NGC 309 (Sc) and NGC 718 (Sa) are almost identical and both
galaxies belong to the same $\beta$ bin} (see Figs. 1 and 2; both NGC
309 and NGC 718 are illustrated in Fig. 2).

\begin{figure}
\vskip500pt
\caption{\small Pitch angle classes in the near-infrared regime, as derived
  from Fourier spectra (Block \& Puerari 1999).  The top row shows two
  tightly wrapped class $\alpha$ spirals (M 83 = NGC5236 and NGC 3223);
  the middle row shows two intermediate class $\beta$ types (NGC 5248
  and NGC 4062), while NGC 5054 and NGC 5921 (bottom row) belong to the
  very open $\gamma$ spiral arm bin. Inverse Fourier transform contours
  are overlayed. Adapted from Block et al. (1999) and Puerari et
  al. (2000).}
\end{figure}

\begin{figure}
\vskip450pt
\caption{\small Spiral galaxies in the dust penetrated regime are binned
 according to three quantitative criteria: H$m$, where $m$ is the
 dominant Fourier harmonic (illustrated here are the two-armed H2
 family); the pitch angle families $\alpha$, $\beta$ or $\gamma$ and
 thirdly the bar strength, derived from the gravitational torque (not
 ellipticity) of the bar. Early type b spirals (NGC 3992, NGC 2543, NGC
 7083, NGC 5371 and NGC 1365) are distributed within all three families
 ($\alpha$, $\beta$ and $\gamma$). Hubble type and dust penetrated
 class are uncorrelated.}
\end{figure}

\section{Bar Strengths Derived from Gravitational Torques}

The most elegant way of measuring bar strength is based on a
definition given by Combes \& Sanders (1981). The methodology uses the
gravitational potential $\Phi(R,\theta)$ of the bar embedded in a disk
and they define the bar strength at radius $R$ as

\begin{equation}
Q_T(R) = {F_T^{max}(R)\over <F_{R}(R)>}
\end{equation}

\noindent
where $F_T^{max}(R)$ =
$({\partial\Phi(R,\theta)/\partial\theta})_{max}$ represents the
maximum amplitude of the tangential force at radius $R$ and
$<F_{R}(R)> = R (d\Phi_0/dR)$ is the mean axisymmetric radial force at
the same radius, derived from the $m$ = 0 component of the
gravitational potential. Let $Q_{bi}$ be the value of $Q_T^{max}$ in
quadrant $i$. Then we define the bar strength as

\begin{equation}
Q_b = \sum_{i=1}^4 Q_{bi}/4.
\end{equation}

Seven gravitational bar strength classes, from class 0 (no bar) to 6
(strong\-est), are elucidated by Buta \& Block (2001).  Galaxies
classified as `SB' in Fig. 3 span a wide range of bar strengths, from
class 2 to class 6.  Apparently `strong' highly elongated bars (e.g. M
83, Fig. 1) may only have a weak gravitational potential bar class
(Fig. 2).

\section{A Dust Penetrated Tuning Fork}

Galaxies are firstly binned according to the dominant Fourier harmonic
H$m$; all galaxies illustrated in Fig. 2 have a regular two-armed
($m=2$) morphology. Higher order harmonics H3 (e.g. NGC 5054, Fig. 1)
and H4 are recognised, but rarer.  A ubiquity of $m=1$ and $m=2$
spirals in the near-infrared is found (Block et al. 1994).  Galaxies
are further subdivided into three dust penetrated groups ($\alpha$,
$\beta$ and $\gamma$) based on the pitch angle of the arms derived
from the Fourier spectra (Block \& Puerari 1999).  These classes are
inextricably related to the {\it rate of shear} in the stellar disk,
as determined by $A/\omega$ (Block et al. 1999; Fuchs, this
volume). Here $A$ is the first Oort constant and $\omega$ is the
angular velocity.  The final quantitative morphological parameter is
the bar strength.

For the galaxies illustrated in Fig. 2, the bar strengths as deduced
from their gravitational potentials, are 1 (NGC 2857), 2 (M 83), 4
(NGC 3992), 1 (NGC 309), 2 (NGC 718), 3 (NGC 2543), 0 (NGC 7083), 2
(NGC 5371) and 5 (NGC 1365).  M 83 is thus fully classified as
H2$\alpha$2. H2= 2-armed spiral; `$\alpha$: tightly wrapped arms and
the final number indicating a bar strength here of 2.  NGC 1365, with
two wide open arms, is classified as H2$\gamma$5.

\begin{figure}
\vskip490pt
\caption{\small The threshold for calling a galaxy `SB' is $\sim$ bar
 torque class 2 (Q$_{b}$ ranges from 0.15 to 0.249 for class 2). The
 Hubble-de Vaucouleurs classifications do not make any further
 discrimination on bar strength Q$_{b}$ (see Eq. 2) beyond this
 threshold. We find that the bars with the strongest gravitational
 torques reach a bar class of 6, where the maximum tangential force
 reaches about 60\% of the mean radial force. Galaxies classified as
 `SA' or `SAB' mostly range from classes 0 to 2. Further details in
 Buta \& Block (2001).}
\end{figure}

\begin{figure}
\vskip450pt
\caption{\small NGC 922 (imaged here in blue light) bears a striking
  resemblance to objects such as HDF2-86 ($z=0.749$) in the
  HDF. Although NGC 922 falls outside traditional classification
  systems, it can readily be classified in the near-infrared template in
  Fig. 2. Dust penetrated imaging shows that a simple two-armed spiral
  (betraying the signature of arm modulation found in several grand
  design prototypes such as Messier 81) is largely responsible for the
  stellar backbone of NGC 922. The bisymmetric spiral ($m=2$) in NGC 922
  is classified as H2$\gamma$ and the complete morphological
  classification would be H2$\gamma$2 since its gravitational bar
  strength is 2. For further details, see Block et al. (2001).}
\end{figure}

\begin{figure}
\vskip450pt
\caption{ Upper left: Groundbased $K'$ image of NGC 922, secured by
  Alan Stockton at Mauna Kea. Upper right: $K'$ image with $m=1$ and $m=2$
  Fourier modes overlayed, as determined from the inverse Fourier
  transform. Middle left: Simulated 3600s image of NGC 922, moved out
  in redshift space to $z=0.7$ ($L$ band) using an 8-m NGST and the
  prescription of Takamiya (1999). Middle right:
  $L$-band image, with $m=1$ and $m=2$ contours overlayed. Bottom right and
  left shows the galaxy redshifted to $z=1.2$ (M band) with and without
  inverse Fourier transform contours.  The $M$-band postage stamp FITS
  images are only 3$''$ on a side, but Fourier spectra can easily
  be generated from them. Adapted from Block et al. (2001).}
\end{figure}

\section{Dust Penetrated Morphology in the High-redshift Universe:
  Clues from NGC 922}

The tuning fork in Fig. 2 may serve as a z=0 template when galaxies at
$z\sim0.5-1$ are imaged {\it in their dust penetrated 2$\mu m$
regime}.  Results from the Hubble Deep Field (HDF) North and South
show a large percentage of high-redshift galaxies whose appearance
falls outside traditional classification systems. The nature of these
objects is poorly understood, but sub-mm observations indicate that at
least some of these systems are heavily obscured (Sanders 1999). This
raises the intriguing possibility that a physically meaningful
classification system for high-redshift galaxies might be more easily
devised at rest-frame infrared wavelengths, rather than in the optical
regime. Practical realization of this idea will become possible with
the advent of the Next Generation Space Telescope (NGST).

In order to explore the capability of NGST for undertaking such
science, Block et al. (2001) present NASA-IRTF and SCUBA observations
of NGC 922, a chaotic system in our local Universe which bears a
striking resemblance to objects such as HDF 2-86 ($z=0.749$) in the
HDF North.  If objects such as NGC 922 are common at high-redshifts,
then this galaxy may serve as a local morphological `Rosetta stone'
bridging low and high-redshift populations. Block et al. (2001) show
that quantitative measures of galactic structure are recoverable in
the rest-frame infrared for NGC 922 seen at high redshifts using NGST,
by simulating the appearance of this galaxy at redshifts $z=0.7$ and
$z=1.2$ in rest-frame $K'$ (Fig. 5).  {\em Our results suggest that the
capability of efficiently exploring the rest-wavelength IR morphology
of high-$z$ galaxies should probably be a key factor in deciding the
final choice of instruments for the NGST}.

\acknowledgements
IP (INAOE, Mexico) and RJB (University of Alabama,
AL) are Distinguished Visitors,
Anglo American Astronomy Program.  Much gratitude to sponsors SASOL
and Anglo-American, to all our collaborators, and to D. A. Crocker for
preparation of Fig. 2.  OSU Galaxy Survey Images NGC 4062/NGC 5054
courtesy J. Frogel (NSF grants AST-9217716 and AST-9617006); M 83 and
NGC 1365 at $Ks$ courtesy O. K. Park and K. Freeman. We dedicate this
paper to the pioneering photographic work of James D. Wray.

\end{document}